# Sky-blue Solution-processed TADF and Hyperfluorescent OLEDs via a High-Work-Function PEDOT:PSS:PFI Hole Injection Layer

Džiugas Litvinas [1], Francisco Tenopala-Carmona [1], Mahni Fatahi [2], Anna Popczyk [1], Eli Zysman-Colman* [2], Malte C. Gather* [1,3]

[1] *Humboldt Centre for Nano- and Biophotonics, Department of Chemistry and Biochemistry, University of Cologne, 50939 Köln, Germany. E-mail: malte.gather@uni-koeln.de*

[2] *Organic Semiconductor Centre, EaStCHEM School of Chemistry, University of St Andrews, St Andrews, UK, KY16 9ST. E-mail: eli.zysman-colman@st-andrews.ac.uk*

[3] *Organic Semiconductor Centre, SUPA School of Physics and Astronomy, University of St Andrews, St Andrews, KY16 9SS, UK.*

## Abstract

Solution-processed OLEDs offer a scalable route to large-area devices, but achieving competitive efficiency in multilayer stacks remains a challenge due to difficulties in hole injection as well as solvent orthogonality between layers. A hole injection layer (HIL) based on a blend of PEDOT:PSS and a perfluorinated ionomer (PFI) addresses both challenges simultaneously: surface segregation of PFI during film drying creates a high-work-function interface improving hole injection into high-ionisation potential host materials, while the insolubility of the blend in aromatic solvents prevents intermixing with subsequently deposited layers. Using this HIL with a solution-processed mCPCN:SpiroAC-TRZ emissive layer, sky-blue OLEDs with a peak external quantum efficiency ($EQE_{max}$) of 21.7%, an external quantum efficiency (EQE) of 20.2% at 100 cd m$^{-2}$, a peak luminance of 29,000 cd m$^{-2}$ and a full width at half maximum (FWHM) emission linewidth of 69 nm are demonstrated. The PFI-based HIL approach is extended to the photochemically stable TADF emitter 5TCzBN ($EQE_{max}$ 18.4%, FWHM 73 nm) and to sensitized / hyperfluorescent OLEDs incorporating the multi-resonance TADF terminal emitters ν-DABNA ($EQE_{max}$ 22.8%, FWHM 26 nm) and the dendrimer TCzBN-BMes ($EQE_{max}$ 20.4%, FWHM 29 nm). These results establish the PEDOT:PSS:PFI HIL as a broadly compatible platform for high-performance solution-processed TADF OLEDs.

## Introduction

Organic light-emitting diodes (OLEDs) have become essential components of high-end consumer displays, and the potential to fabricate them by solution processing — ideally using roll-to-roll processing compatible with large-area substrates — has been a central promise of organic electronics since its early days. Solution processing offers a compelling alternative to the thermal evaporation methods currently dominant in industry, which require ultra-high vacuum conditions and are associated with significant material consumption [1], [2]. However, translating this promise into devices that meet industry standards for efficiency, stability, and colour purity requires careful navigation of both material and processing constraints — not least because achieving competitive performance generally demands a multilayer architecture in which charge injection, transport, and recombination are each handled by a dedicated material.

A central challenge for solution processing such multilayer stacks is maintaining the structural integrity of each layer during the deposition of subsequent ones. To achieve this, each layer should ideally be deposited from a solvent orthogonal to those used below — i.e., one that does not dissolve or swell the preceding layers. In addition, each material/solvent combination must offer sufficient solubility and form smooth, highly amorphous films — properties that are far from guaranteed for an arbitrary combination of materials and solvents. This requirement means that in practice, solvent orthogonality is often compromised. A common example of this tension is the combination of PEDOT:PSS as a hole injection layer (HIL) with an emission layer (EML) deposited directly from a non-polar solvent such as chlorobenzene or toluene — a pairing that is naturally orthogonal since PEDOT:PSS is processed from an aqueous dispersion. However, direct hole injection from PEDOT:PSS into the EML is often impeded by a substantial energy barrier to the highest occupied molecular orbital (HOMO) of typical organic hosts, a problem that becomes particularly acute for blue emissive thermally activated delayed fluorescence (TADF) emitters and the thus required high triplet energy host materials. An interlayer of, for example, poly(N-vinylcarbazole) (PVK), potentially doped with small molecule hole-transport materials like TCTA, is therefore often introduced to bridge this energy offset [3]. PVK is itself typically deposited from chlorobenzene, and deposition of the EML can therefore partially dissolve it, causing layer intermixing that can introduce bulk trapping sites and exciton quenching centres in the EML and lead to inconsistent device performance [4], [5], [6]. Despite these challenges, solution-processed TADF OLEDs have demonstrated maximum EQEs ($EQE_{max}$) above 20% [1], [7], [8], [9], [10], [11], [12], [13], [14] though performance often lags behind vacuum-deposited equivalents [15], [16], [17].

An alternative approach that raises the work function of PEDOT:PSS without introducing a separate interlayer is to blend it with a perfluorinated ionomer (PFI). Due to the hydrophobicity of PFI, it preferentially segregates to the top surface of the film during drying [18], creating a vertical gradient in ionisation energy from approximately −5.1 eV near the substrate to −5.8 eV at the surface [18], [19]. This effect has been exploited to improve the performance of perovskite solar cells and light-emitting diodes [20], [21]. In the context of OLEDs, Sachnik *et al.* demonstrated that blending PEDOT:PSS with PFI enables ohmic hole injection into vacuum-deposited EML consisting of high ionisation energy hosts and TADF emitters and that this translated into more efficient device operation than in the device without the PFI [19]. However, whether the same HIL strategy can support solution-processed EMLs without layer intermixing — and thereby eliminate a vacuum processing step in OLED fabrication — remains an open question.

Here, we demonstrate that a HIL based on a blend of PEDOT:PSS and PFI can indeed support solution-processed EMLs and use this as the basis for efficient sky-blue TADF OLEDs that outperform common reference architectures in terms of efficiency. The PEDOT:PSS:PFI HIL enables direct hole injection into the EML without requiring an interlayer, and we show that its resistance to non-polar solvents allows the EML to be spin-coated directly on top without intermixing. To show the generality of our approach, we demonstrate compatibility with several solution-processed EML configurations: We first evaluate two commercially available TADF emitters in a host-guest EML: SpiroAC-TRZ, which offers high efficiency [22] but limited stability [24], and 5TCzBN, which offers considerably greater stability at the cost of somewhat lower intrinsic efficiency [24], [41]. An optimised device based on SpiroAC-TRZ achieves an $EQE_{max}$ of 21.7%, decreasing to an $EQE_{100}$ of 20.2% at a luminance of 100 cd $m^{-2}$, with a peak wavelength and full width at half maximum (FWHM) of the electroluminescence (EL) spectrum of $\lambda_{EL}$ = 492 nm and FWHM = 69 nm, respectively, and a maximum luminance of $L_{max}$ = 29,000 cd $m^{-2}$. We extended the concept to hyperfluorescent (HF) devices, in which a small quantity of a high radiative-rate multi-resonant TADF (MR-TADF) terminal emitter—here the widely used ν-DABNA [25] or the dendrimer TCzBN-BMes [26] —is added to narrow the emission linewidth and boost device efficiency. Optimised ν-DABNA-based HF OLEDs reach an $EQE_{max}$ of 22.8% ($\lambda_{EL}$ = 472 nm, FWHM = 26 nm). Together, these results establish the PEDOT:PSS:PFI HIL as a simple and scalable route to high-performance solution-processed TADF OLEDs that is broadly compatible with state-of-the-art EML materials.

# Results

## Device architecture

To systematically evaluate the PEDOT:PSS:PFI blend as a HIL for solution-processed OLEDs, we compared three device architectures that share a common layer stack above the HIL, as schematically illustrated in Figure 1. The first two serve as reference architectures: bare PEDOT:PSS (Figure 1a), representing the simplest solution-processed HIL, and a PEDOT:PSS/PVK:TCTA bilayer (Figure 1b), in which a PVK:TCTA (4:1) interlayer is intended to bridge the ionisation energy offset between PEDOT:PSS and the EML. The third architecture (Figure 1c) instead uses a single spin-coated layer of a PEDOT:PSS:PFI blend (1:6:14 by weight). The graded ionization potential profile of the PEDOT:PSS:PFI layer enables direct hole injection into high ionisation potential EML materials without requiring a discrete interlayer, while its resistance to non-polar solvents allows the EML to be spin-coated directly on top without risking intermixing.

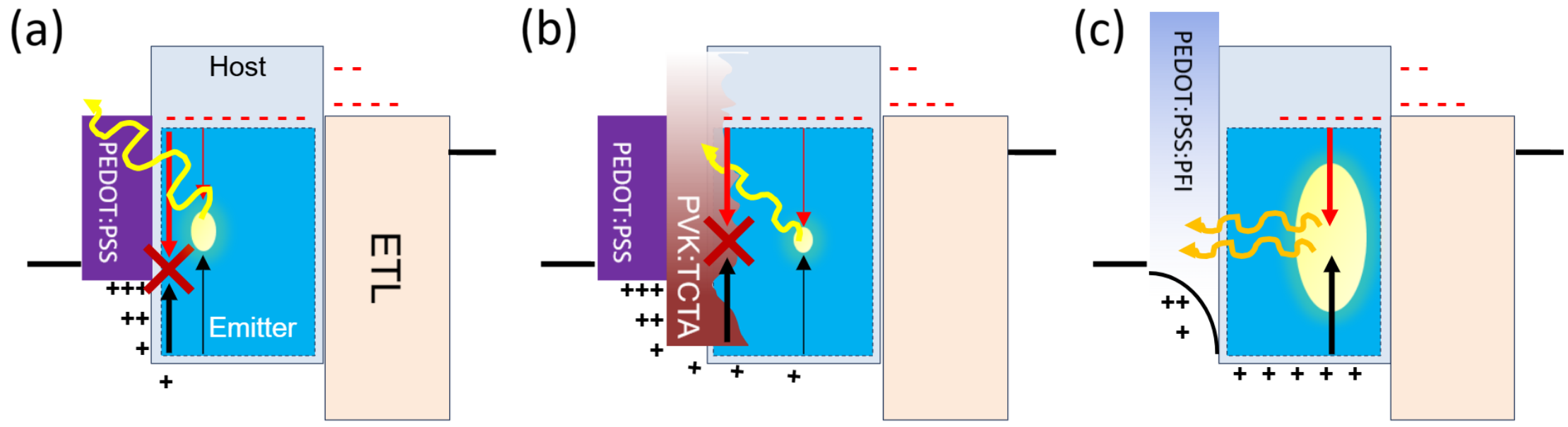

Figure **1**. Schematic comparison of the three HIL architectures investigated in this work. (a) The large energy offset between the work function of a bare PEDOT:PSS HIL and the host HOMO can lead to accumulation of positive charge at the HIL/EML interface, impeding hole injection into the EML. (b) The addition of a PVK:TCTA interlayer can reduce the injection barrier, but its partial dissolution by the EML solvent likely introduces inhomogeneity and intermixing at the PVK:TCTA/EML interface, creating exciton quenching centres. (c) In a PEDOT:PSS:PFI blend surface segregation of PFI during film formation produces a smooth, high-work-function surface that is likely to promote hole injection into the host, suppress interfacial intermixing, and reduce exciton–polaron quenching at the HIL/EML interface.

For the EML, we selected the high triplet energy bipolar host mCPCN [27] ($E_{T1}$ = 3.03 eV, chemical structure shown in Figure S3a), which has a higher electron than hole mobility ($1.1 \times 10^{-5}$ vs. $1.9 \times 10^{-6}$ $cm^2$ $V^{-1}$ $s^{-1}$), and doped it with the TADF emitter SpiroAC-TRZ [22] ($E_{T1}$ = 2.51 eV, Figure S3b). The deeper LUMO of SpiroAC-TRZ (-3.1 eV) relative to that of mCPCN (-2.24 eV) means it acts as an electron trap within the host matrix, thus making charge balance in the device a function of emitter doping concentration. SpiroAC-TRZ is characterised by fast radiative and RISC rates and low aggregation-caused emission quenching, even at high doping concentrations, making it well suited to the relatively high guest fraction (>10 wt%) required to achieve charge balance in an mCPCN host. Both mCPCN and SpiroAC-TRZ readily dissolve in chlorobenzene, whose relatively high boiling point (132 °C) makes it suitable for spin-coating of the EML. To demonstrate the generality of the PEDOT:PSS:PFI HIL approach for OLEDs with solution-processed EMLs, we extended the study to devices with the TADF emitter 5TCzBN [24], [41], which offers greater operational stability than devices with SpiroAC-TRZ, and to HF-device configurations employing the narrow-linewidth MR-TADF terminal emitters $\nu$-DABNA (Figure S3d) [25] and TCzBN-BMes (Figure S3e) [26].

## Optimisation of solution-processed layers

Prior to device fabrication, the optical properties of the PEDOT:PSS:PFI blend were characterised by variable-angle spectroscopic ellipsometry (VASE) [28]. Films were prepared on both glass and silicon substrates, and the measured data were analysed using physical oscillator models to extract optical constant spectra, i.e., the complex refractive index as function of wavelength (Figure 2a, b). The spectrum of the blend was reproduced by optimising the mixing ratio of the individually determined dielectric functions of PEDOT:PSS and PFI, and the best fit was obtained for the weight ratio used in the precursor.

A critical requirement for solution-processed multilayer stacks is solvent orthogonality: each deposited layer must withstand exposure to solvents used in subsequent coating steps. Profilometry of individually spin-coated films before and after rinsing with chlorobenzene (Figure S1) confirmed that both PEDOT:PSS and PFI films are unaffected by chlorobenzene exposure, establishing that the PEDOT:PSS:PFI HIL is likewise stable to chlorobenzene. A film of PVK:TCTA, by contrast, was almost completely dissolved under the same conditions, underscoring the intermixing risk inherent to bilayer architectures.

A practical challenge arising from the PFI content of the blended HIL is its poor wettability towards the aromatic solvents used for EML spin-coating. This was addressed by exposing the annealed HIL to a brief low-pressure radio-frequency (RF) $O_2$ plasma treatment prior to EML deposition, which is assumed to introduce polar surface groups that enhance solvent wettability (Figure 2c). As shown in Figure 2d and e, the contact angle drops from approximately 30° to 5° following treatment, confirming effective surface activation. Atomic force microscopy (AFM) topographic imaging shows that the surface root mean square roughness remains similar after plasma treatment (before: 3.2 nm, after: 3.5 nm; Figure S2), confirming that the bulk film morphology is not significantly disrupted. However, the overall film thickness reduced by 25 nm during plasma treatment, indicating that the upper PFI-rich region of the PEDOT:PSS:PFI film was likely removed, at least in part. Furthermore, plasma treatment might introduce defects that can accelerate device degradation and the loss in thickness had to be compensated by starting with a correspondingly thicker initial layer. Therefore, plasma treatment was only used for the first set of OLEDs (Figure 3c, see below). Following optimization of the spin-coating protocol, the EML was spin-coated directly on the untreated PEDOT:PSS:PFI layer for the later sets of devices.

With the HIL processing conditions established, the photophysical properties of the mCPCN:SpiroAC-TRZ EML were characterised as a function of the emitter doping concentration. Photoluminescence quantum yield (PLQY) measurements (Figure S5a) reveal two concentration-dependent trends: increasing SpiroAC-TRZ content produces a progressive spectral red-shift accompanied by a decrease in PLQY. This is indicative of aggregation-caused quenching. Additionally, the polar mCPCN host ($\mu_{\mathrm{GS}}$ = 6.19 D) introduces a solvation-induced red-shift, which is characteristic of charge-transfer emitters in polar environments [27]. Despite these effects, the PLQY remains high across a wide range of concentrations, reaching 97% at 40 wt% of SpiroAC-TRZ, which is the concentration selected for subsequent devices.

Finally, the orientation of the SpiroAC-TRZ transition dipole moment (TDM) in the solution-processed mCPCN:SpiroAC-TRZ (40 wt%) EML was assessed by angle-resolved photoluminescence (ARPL). The anisotropy parameter was determined to be $a$ = 0.31 ± 0.02 (Figure S5b), indicating an effectively isotropic emitter orientation. This contrasts with vacuum-deposited SpiroAC-TRZ films in the same host, which exhibit a strongly horizontal dipole ratio of 83% ($a$ = 0.17) [22], and is consistent with the general observation that many solution-processed doped films of small-molecule emitters tend towards isotropic orientation, irrespectively of host, due to the absence of the directed intermolecular interactions that drive alignment during vacuum deposition [29], [30], [31], [32], [33], [34], [35]. Notable exceptions include oligomers [36], dendrimers [9], compatible geometry host-bulky emitter systems [11] and liquid-crystal molecules [37], [38].

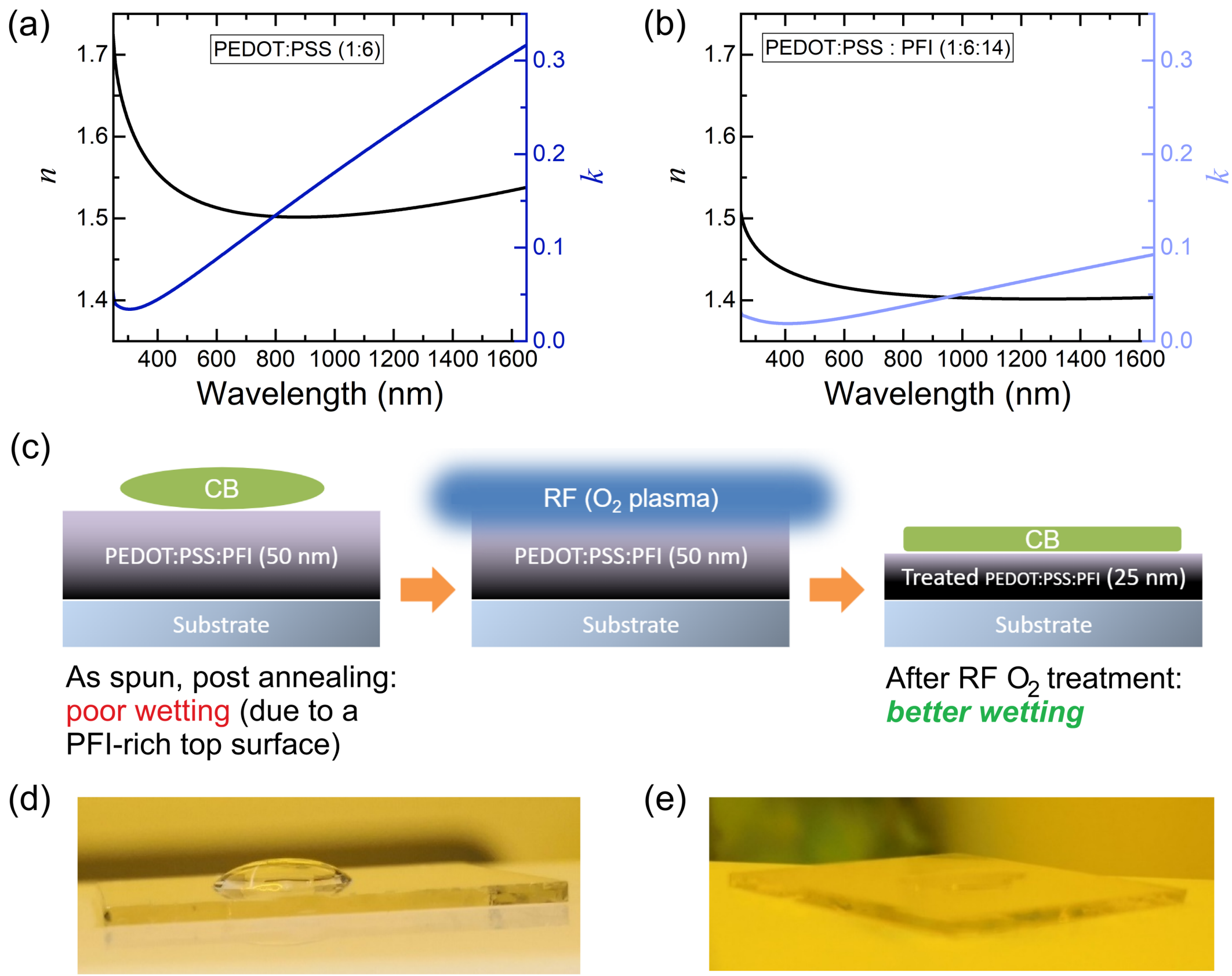

Figure **2**. Optical constant spectra of spin-coated films, as determined from ellipsometry data for (a) PEDOT:PSS (1:6) and (b) PEDOT:PSS:PFI (1:6:14). (c) Scheme for optional RF treatment of PEDOT:PSS:PFI layers. (d) As cast 50 nm film with poor solvent wettability. (e) After 20 s of low pressure $O_2$ RF plasma treatment, film thickness decreased to 25 nm and wettability was substantially increased.

## Effect of hole injection layer on SpiroAC-TRZ device performance

Having established the optical and physical properties of the solution-processed HIL and EML, we fabricated OLEDs based on each of the three architectures described above and compared their performance. Figure 3a–c summarizes the layer thicknesses and compositions for the different devices and shows the energy level diagrams, illustrating the different charge injection pathways for each architecture. The *J*–*V*–*L* and EQE characteristics of the best-performing device for each architecture (see below for details) are shown in Figure 3d and e, and the key performance metrics are summarised in Table 1.

All three devices have similar turn-on voltages ($V_{on}$ = 4.0–4.3 V at 10 cd $m^{-2}$), indicating that the choice of HIL does not significantly impede initial charge injection under forward bias. The minor differences in the EL spectra between devices reflect differences in optical cavity length arising from the different HIL thicknesses. The bare PEDOT:PSS device reaches higher current densities and luminance than the PEDOT:PSS:PFI device across the measured voltage range; we attribute this to a lower electrical conductivity of the blended HIL, in which the addition of insulating PFI chains likely reduces the overall conductivity relative to neat PEDOT:PSS.

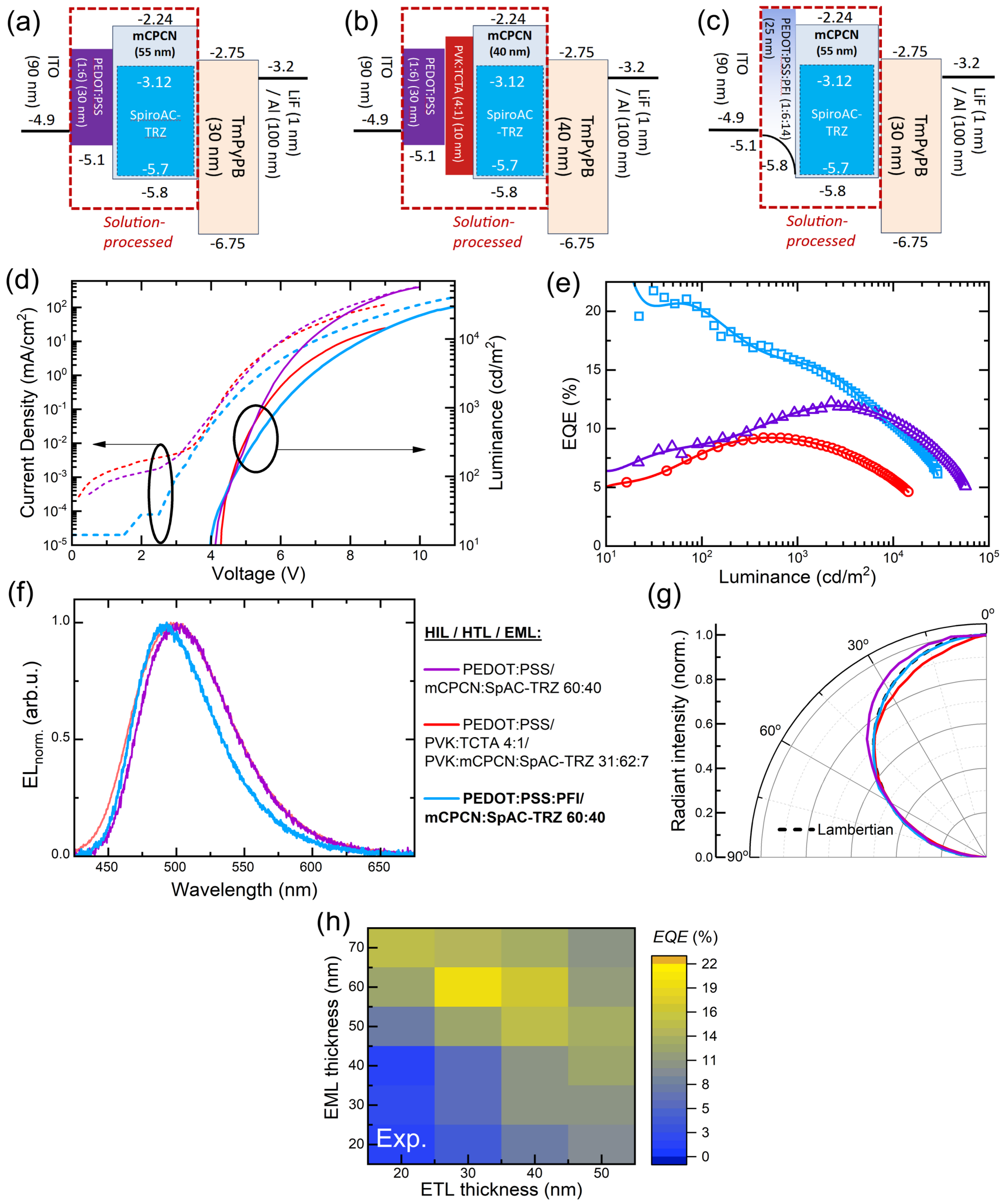

Figure 3. Solution-processed OLEDs based on a mCPCN:SpiroAC-TRZ (40 wt%) EML for three HIL architectures: bare PEDOT:PSS, PEDOT:PSS/PVK:TCTA, and PEDOT:PSS:PFI. (a–c) Assumed energy level diagrams summarising the layer compositions and thicknesses for each architecture. (d) Current density–voltage–luminance, (e) EQE–luminance (symbols: data, solid lines: guides to the eye), (f) electroluminescence spectra, and (g) normalised radiant intensity as a function of viewing angle for the best-performing device from each architecture; the dashed line indicates a Lambertian emission profile. (h) $EQE_{max}$ as a function of EML and ETL thickness for the PEDOT:PSS:PFI architecture.

However, the PEDOT:PSS:PFI device performs substantially better than the devices using either of the reference architectures in terms of EQE, achieving an $EQE_{max}$ of 21.7%, compared to 12.2% for the device with the bare PEDOT:PSS and 9.2% for the device with the PEDOT:PSS/PVK:TCTA (4:1). We attribute the superior efficiency of the PEDOT:PSS:PFI device primarily to the reduction of the hole injection barrier. The graded ionisation potential profile of the PFI-rich surface enables direct hole injection into the HOMO of the mCPCN host, likely reducing the accumulation of interfacial charges and associated exciton–polaron quenching that may limit the efficiency of the bare PEDOT:PSS device. The improved efficiency relative to the PEDOT:PSS/PVK:TCTA bilayer device is consistent with the absence of intermixing between the PVK:TCTA interlayer and the EML; such intermixing likely introduces bulk trapping sites that quench excitons and compromise batch-to-batch reproducibility. In addition, the low electron mobility of PVK and its reduced hole mobility (~$10^{-7}$ cm$^2$ V$^{-1}$ s$^{-1}$) [3] compared to mCPCN (1.9 × $10^{-6}$ cm$^2$ V$^{-1}$ s$^{-1}$) [39] likely lead to charge imbalance when layer intermixing occurs.

The PEDOT:PSS:PFI device achieved an $EQE_{100}$ of 20.2% and $EQE_{1000}$ of 15.8%. A peak luminance of 29,000 cd m$^{-2}$ is reached before the onset of significant efficiency roll-off and device degradation. The EL spectrum peaks at 492 nm (FWHM = 69 nm, Figure 3f), coinciding with the PL emission peak of an mCPCN:SpiroAC-TRZ film (40 wt%, Figure S5), indicating that the device cavity is well-tuned for outcoupling at the emission wavelength. All three devices exhibit near-Lambertian angular emission profiles (Figure 3g).

The optimal combination of EML and ETL thickness was identified by analysing the $EQE_{max}$ values for a series of devices that had independently varying thicknesses of both layers (Figure 3h). This reveals that peak efficiency is achieved for an EML thickness of approximately 55 nm and an ETL thickness of 30 nm. For EML thicknesses above 65 nm device efficiency drops sharply, which we attribute to a probable increase in the number of pinholes in thicker spin-coated films, where slower solvent drying kinetics may compromise film uniformity and introduce current shunting pathways. The observed dependence of device efficiency on ETL thickness likely reflects a combination of optical and electrical effects: first, the ETL thickness modulates the position of the recombination zone relative to the optical field maximum of the OLED cavity and thereby the outcoupling efficiency; second, through its influence on electron transport, the ETL thickness affects the balance of electron and hole currents within the EML.

Table **1**. Performance metrics of solution-processed OLEDs based on a mCPCN:SpiroAC-TRZ (40 wt%) EML for the three HIL architectures investigated in this work.

| HIL (/HTL) | $V_{on}$ (V) | $EQE_{max/100/1000/Lmax}$ (%): | $L_{max}$ ($10^3$ cd m$^{-2}$) | $\lambda_{EL}$ (nm) | FWHM (nm) | CIE (x, y) |
|---|---|---|---|---|---|---|
| PEDOT:PSS | 4.2 | 12.2 / 8.7 / 11.4 / 5.1 | 57.1 | 501 | 77 | 0.203, 0.479 |
| PEDOT:PSS/PVK:TCTA | 4.3 | 9.2 / 7.8 / 9.0 / 4.6 | 14.4 | 497 | 84 | 0.186, 0.409 |
| PEDOT:PSS:PFI | 4.0 | 21.7 / 20.2 / 15.8 / 6.1 | 29.0 | 492 | 69 | 0.175, 0.419 |

$V_{on}$ is defined as the voltage at 10 cd m$^{-2}$. EQE values are reported as peak EQE, EQE at the luminance levels indicated, and EQE at the peak luminance $L_{max}$ achieved by each device.

## Extension of the PEDOT:PSS:PFI approach to a stable TADF emitter

Having demonstrated efficient sky-blue OLEDs with SpiroAC-TRZ as the emitter, we next evaluated the PEDOT:PSS:PFI HIL in combination with the TADF emitter 5TCzBN (Figure 4a) [24], [41]. This serves to assess whether the HIL strategy is transferable to a more intrinsically stable emitter system. 5TCzBN possesses the highest difference between the C–N bond dissociation energy (*BDE*) and $E_{T1}$ (*BDE* – $E_{T1}$ = 1.96 eV) in a recently surveyed dataset of 35 blue TADF emitters [24]. It was therefore identified as the most intrinsically photochemically stable emitter in that set, a property attributed to strong steric shielding of the labile C–N bonds by the *tert*-butyl substituents on the carbazole donors [41]. Relative to SpiroAC-TRZ, however, 5TCzBN has a somewhat slower radiative rate constant ($k_r$ of $1.59 \times 10^7$ s$^{-1}$ [41] *vs.* $5.9 \times 10^7$ s$^{-1}$ [22]), larger singlet–triplet energy gap ($\Delta E_{ST}$ of 170 meV vs 72 meV), and lower PLQY (85–87% *vs.* near 100%), which may limit device efficiency. Selecting 5TCzBN, therefore, allows for an evaluation of whether the efficiency penalty associated with a lower PLQY and larger $\Delta E_{ST}$ can be mitigated by the PEDOT:PSS:PFI HIL strategy.

The stack architecture and energy diagram of the 5TCzBN-based device are shown in Figure 4b. For these devices, the EML composition consisted of 30 wt% 5TCzBN doped in mCPCN, with the ETL thickness being varied from 25 to 40 nm in 5-nm steps. The best-performing device achieved an $EQE_{max}$ of 18.4% and an $EQE_{100}$ of 15.8% and reaches a peak luminance of 5,370 cd m$^{-2}$ before the onset of significant efficiency roll-off and degradation (Figure 4c,d). The $\lambda_{EL}$ is 489 nm (FWHM = 73 nm, Figure 4e), which is close to that of the SpiroAC-TRZ device. Compared to the SpiroAC-TRZ devices, the $EQE_{max}$ is lower and the efficiency roll-off is more pronounced. Further, the angular emission profiles are super-Lambertian across the ETL thickness series (Figure 4f). We tentatively attribute these observations to a combination of the lower PLQY and the shallower LUMO of 5TCzBN (−2.7 eV compared to −3.1 eV for SpiroAC-TRZ), which likely makes electron injection from TmPyPB more difficult, and may also affect charge balance and shift the recombination zone towards the cathode side of the EML.

Nevertheless, $EQE_{max}$ and the CIE colour point of the devices (0.209, 0.410) are comparable to the performance of the original literature reported vacuum-deposited 5TCzBN-based OLEDs, which achieved an $EQE_{max}$ of 21.2%, a $\lambda_{EL}$ of 490 nm, and a CIE colour point of (0.21, 0.41) [41]. This is particularly notable given that our device is processed largely from solution. More recently, vacuum-deposited devices employing a perdeuterated variant of 5TCzBN in an exciplex host system have reached an $EQE_{max}$ of 26.0% and an $EQE_{1000}$ of 22.7% [23], illustrating the performance headroom that remains for solution-processed architectures. Nonetheless, our results demonstrate that the PEDOT:PSS:PFI HIL strategy is broadly compatible with 5TCzBN and that solution processing of this emitter is viable without a significant loss of efficiency relative to vacuum-deposited references.

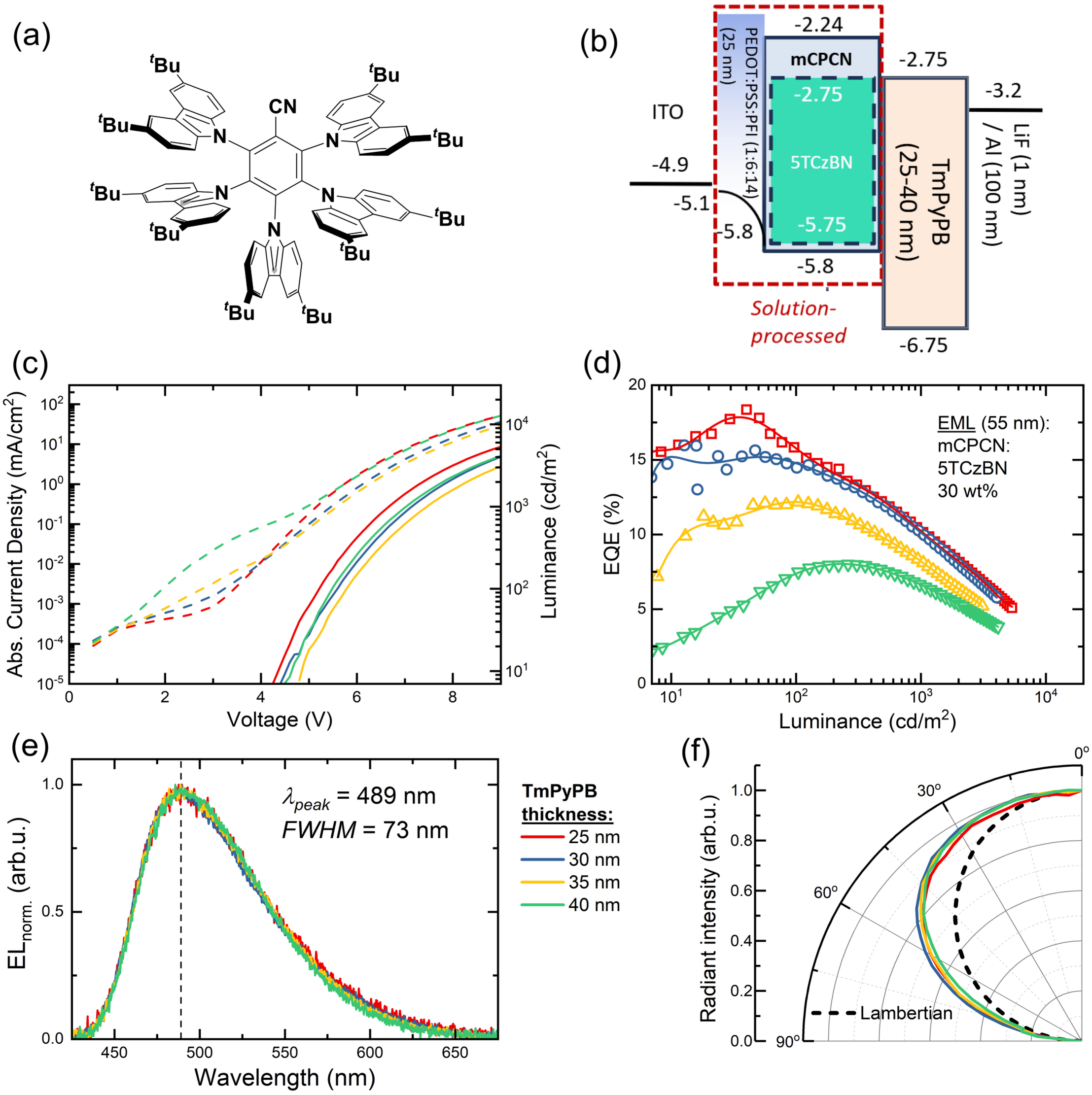

Figure **4.** Characteristics of solution-processed OLEDs based on a mCPCN:5TCzBN (30 wt%) EML with the PEDOT:PSS:PFI HIL. (a) Molecular structure of 5TCzBN. (b) Assumed energy level diagram of the device stack. (c) Current density–voltage–luminance and (d) EQE–luminance characteristics for devices with ETL thicknesses of 25, 30, 35, and 40 nm (symbols: data, solid lines: guides to the eye). (e) Electroluminescence spectra for the same series of devices. (f) Normalised radiant intensity as a function of viewing angle; the dashed line indicates a Lambertian emission profile.

## Sensitized devices with MR-TADF terminal emitters

MR-TADF emitters have attracted considerable interest as terminal emitters in HF OLEDs owing to their rigid polycyclic aromatic framework, which suppresses excited-state structural relaxation and vibronic coupling, resulting in bright, narrowband emission [25], [42], [43]. By contrast, the intramolecular long-range charge-transfer character (LRCT) of donor–acceptor TADF emitters gives rise to much broader emission spectra and hence a reduced colour gamut (e.g., the EL spectra of SpiroAC-TRZ and 5TCzBN have FWHM of 69 and 73 nm, respectively, in our devices). In HF OLEDs, electrical excitation generates excitons primarily on the host and sensitizer, which are then transferred to the MR-TADF terminal emitter via Förster resonance energy transfer (FRET), thereby combining efficient triplet harvesting of a TADF sensitizer with narrow, high-radiative-rate emission of an MR-TADF acceptor. Here we employ this strategy using 5TCzBN as the sensitizer — building directly on the devices described in the previous section — in combination with two terminal emitters: the benchmark MR-TADF emitter ν-DABNA [25] and the MR-TADF compound TCzBN-BMes, which was custom-designed for solution-processability. The PL spectrum of the 5TCzBN sensitizer and the absorption and PL spectra of the terminal emitters ν-DABNA and TCzBN-BMes are compared in Figure 5a,b, with the chemical structures of both terminal emitters are shown in Figure S3.

To assess the suitability of these terminal emitter-sensitizer pairs prior to device fabrication, we estimated the FRET efficiency for several sensitizer concentrations. Our analysis (described in the Supporting Information section *FRET efficiency calculation*) shows that at a 30 wt% sensitizer loading, the energy transfer efficiency already reaches 86% for both terminal emitters at a concentration of 1 wt% (Table 2), fully sufficient for realizing sensitized devices [44]. OLEDs were therefore fabricated with an EML composition of mCPCN:5TCzBN:terminal emitter (69:30:1 wt%, Figure 5c,d). Both HF devices outperform the sensitizer-only 5TCzBN device in terms of $EQE_{max}$, efficiency roll-off, and peak luminance (Figure 5e,f). Notably, their $EQE_{max}$ values exceed even those of the mCPCN:SpiroAC-TRZ device. The ν-DABNA-based device achieved an $EQE_{max}$ of 22.8% at $\lambda_{EL}$ of 472 nm (FWHM = 26 nm) and CIE colour point of (0.169, 0.329), while the TCzBN-BMes-based device reached an $EQE_{max}$ of 20.4% and emitted at $\lambda_{EL}$ of 477 nm (FWHM = 29 nm), with a CIE colour point of (0.142, 0.306) (Figure 5g). The substantially narrower FWHM relative to the sensitizer-only devices confirms that emission originates primarily from the MR-TADF terminal emitter in both cases. The improved efficiency roll-off relative to the 5TCzBN sensitizer-only device is consistent with the higher radiative rate of the MR-TADF terminal emitters (Table 2), which more rapidly depopulate the singlet exciton pool and reduce the probability of bimolecular quenching processes. The angular emission profiles of both HF devices are closer to Lambertian than the super-Lambertian profiles observed for the sensitizer-only 5TCzBN devices (Figure 5h).

Solution-processed HF OLEDs combining small-molecule TADF sensitizers with MR-TADF terminal emitters remain relatively scarce in the literature, although several groups have recently reported devices with somewhat higher $EQE_{max}$ values than those achieved here, using purpose-designed host–emitter–sensitizer combinations optimised for efficient FRET [1], [9], [10], [11], [13], [45]. Our results demonstrate that competitive HF device performance is achievable without such purpose-designed pairings — using ν-DABNA as the terminal emitter and 5TCzBN as the sensitizer, the latter chosen for its photochemical stability rather than optimised spectral overlap — and that the PEDOT:PSS:PFI HIL strategy is broadly compatible with ternary HF EML systems.

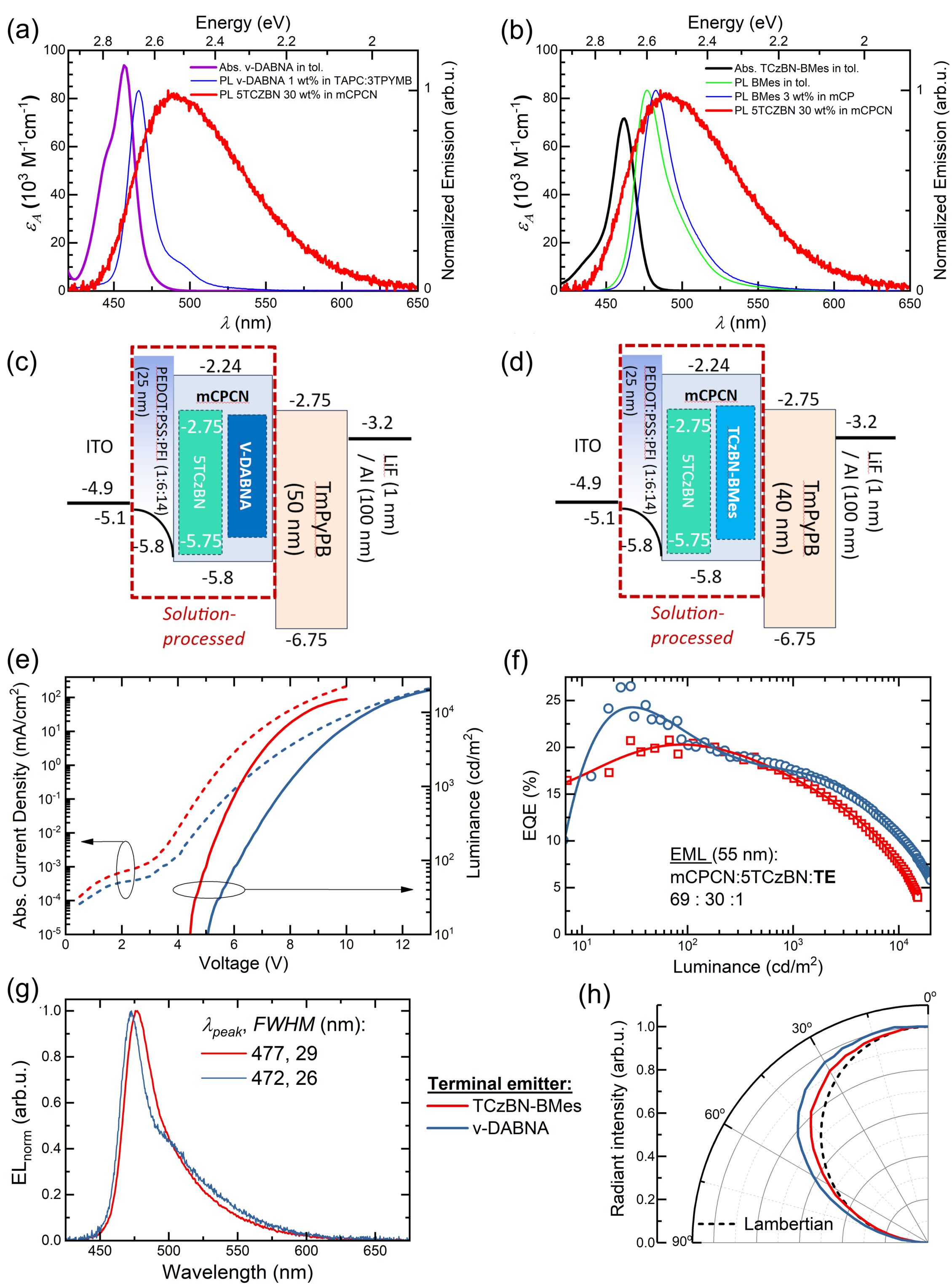

Figure **5.** Characterisation of solution-processed HF OLEDs based on a mCPCN:5TCzBN:terminal emitter (69:30:1 wt%) EML with the PEDOT:PSS:PFI HIL. (a,b) Extinction coefficient and photoluminescence spectra of terminal emitters v-DABNA and TCzBN-BMes together with the photoluminescence spectrum of 5TCzBN, illustrating the spectral overlap relevant to FRET. (c,d) Assumed energy level diagrams for the v-DABNA- and TCzBN-BMes-based devices. (e) Current density–voltage–luminance and (f) EQE–luminance characteristics (symbols: data, solid lines: guides to the eye). (g) Electroluminescence spectra. (h) Normalised radiant intensity as a function of viewing angle; the dashed line indicates a Lambertian emission profile.

Table 2. FRET calculation parameters and device performance metrics for solution-processed HF OLEDs based on a mCPCN:5TCzBN:terminal emitter EML with the PEDOT:PSS:PFI HIL. For details on FRET estimations and definition of all parameters see Supporting Information section "*FRET Efficiency Calculation*".

| | **Rates** | | **FRET calculation** | | | | | | **Device performance** | | | | |
|---|---|---|---|---|---|---|---|---|---|---|---|---|---|
| **EML** | $k_r$ ($10^8$ s$^{-1}$) | $k_{rISC}$ ($10^4$ s$^{-1}$) | $J$ ($10^{-24}$ mol m$^6$) | | $R_{FRET}$ (nm) | | $\eta_{FRET}$ for 5TCzBN fraction in mCPCN | | $V_{on}$ (V) | $\lambda_{EL}$ / *FWHM* (nm) | *EQE*$_{max/100/1000/Lmax}$ (%) | $L_{max}$ (cd m$^{-2}$) | CIE (x, y) |
| | | | 10 wt% | 30 wt% | 10 wt% | 30 wt% | 10 wt% | 30 wt% | | | | | |
| Sensitizer-only | - | - | - | - | - | - | - | - | 4.3 | 489 / 73 | 18.4 / 15.8 / 10.2 / 5.1 | 5,370 | 0.209, 0.410 |
| ν-DABNA | 2.0 | 20 | 3.66 | 4.33 | 3.460 | 3.557 | 36.3% | 85.8% | 5.0 | 472 / 26 | 22.8 / 20.2 / 17.5 / 5.8 | 20,000 | 0.169, 0.329 |
| TCzBN-BMes | 1.65 | 1.03 | 3.69 | 4.35 | 3.464 | 3.560 | 36.4% | 85.9% | 4.4 | 477 / 29 | 20.4 / 20.3 / 16.7 / 4.0 | 15,100 | 0.142, 0.306 |

$V_{on}$ is defined as the voltage at 10 cd m$^{-2}$. The EQE column lists peak EQE and the EQE at the luminance levels indicated.

## Conclusions

We have demonstrated efficient sky-blue solution-processed OLEDs based on a PEDOT:PSS:PFI HIL, achieving an EQE of 20.2% at 100 cd m$^{-2}$ and a peak luminance of 29,000 cd m$^{-2}$ with SpiroAC-TRZ as the TADF emitter. Devices with the PEDOT:PSS:PFI HIL consistently outperformed those with both bare PEDOT:PSS and PEDOT:PSS/PVK reference HILs, which we attribute to a combination of favourable energy level alignment, reduced interfacial polaron–exciton quenching, and suppression of layer intermixing at the HIL/EML interface. We further demonstrated that this HIL strategy is transferable to the photochemically stable emitter 5TCzBN, and that 5TCzBN can serve as an effective sensitizer in solution-processed HF OLEDs incorporating ν-DABNA and TCzBN-BMes as MR-TADF terminal emitters, with the best HF device achieving an $EQE_{max}$ of 22.8% and a FWHM of 26 nm.

The surface segregation of PFI within the PEDOT:PSS:PFI blend does not only form a PFI-rich interlayer at the HIL/EML interface, but it may also play additional protective roles that warrant further investigation in the future. Given the expected high LUMO energy of PFI, this surface layer could act as an electron-blocking interlayer, reducing electron leakage into the HIL and thereby limiting oxidative degradation of PEDOT:PSS under device operation. Furthermore, the presence of a thin, relatively insulating interlayer through which holes cannot be rapidly transported may reduce the surface charge density at the HIL/EML interface, thereby limiting exciton–polaron quenching during device operation. We offer these points as hypotheses for future work, noting that confirming them will require dedicated charge-transport and degradation measurements that are beyond the scope of the present study.

Looking ahead, we anticipate that replacing the relatively more polar mCPCN host with a lower-polarity alternative could further improve both device efficiency and colour purity. Polar host materials, while beneficial for solution processability due to their relatively higher solubility in the relevant solvents, can stabilise charge-transfer excited states through dipolar relaxation, leading to red-shifted and broadened emission spectra and potentially reduced radiative rates. Host materials that are low polarity and have high triplet energy, a wide bandgap, and sufficient solution processability for balanced charge transport therefore represent a promising direction for future optimisation of solution-processed TADF and HF OLEDs. More broadly, the solution-processed OLED stacks developed in this work provide well-characterised reference architectures for future studies requiring a reliable, reproducible baseline for device performance and charge transport in solution-processed organic emitter systems.

## Methods

**Materials**

The commercially available materials used in this work are poly(3,4-ethylenedioxythiophene):polystyrene sulfonate – PEDOT:PSS (1.3-1.7 wt% aqueous dispersion; Al 4083, Ossila); PFI (ca. 21 wt% dispersion in a mixture of alcohol and deionized water; Nafion® D2020, Ion Power); 9-(3-(9H-carbazol-9-yl)phenyl)-9H-carbazole-3-carbonitrile – mCPCN (sublimed >99% purity, Lumtec); spiroacridine-triazine – SpiroAC-TRZ (sublimed >99% purity, Lumtec); 2,3,4,5,6-pentakis(3,6-di-*tert*-butyl-9H-carbazol-9-yl)benzonitrile – 5TCzBN (sublimed >99% purity, Ossila); N7,N7,N13,N13,5,9,11,15-octaphenyl-5,9,11,15-tetrahydro-5,9,11,15-tetraaza-19b,20b-diboradinaphtho[3,2,1-de:1',2',3'-jk]pentacene-7,13-diamine – ν-DABNA (sublimed >99% purity, Lumtec); 1,3,5-tri(*m*-pyridin-3-ylphenyl)benzene – TmPyPB (sublimed >99.8% purity, Lumtec); lithium fluoride – LiF (>99.99% purity, Lumtec); aluminium – Al (99.99% purity, Kurt J. Lesker Co.). TCzBN-BMes was synthesised as described in [26].

**Spin-coating thickness calibration**

Prior to device fabrication, a spin-coating calibration curve was established for the EML thickness as a function of spin speed. Films were prepared by spin-coating the mCPCN:SpiroAC-TRZ solution described in the OLED Fabrication section below onto glass substrates coated with the PEDOT:PSS:PFI HIL, and the resulting bilayer thicknesses were determined by profilometry (Bruker Dektak XT). After subtracting the isolated HIL thickness, the EML thickness values $d$ were plotted against spin-coating speed $\omega$ and fitted to a power law:

$$d = A\omega^{-\gamma}. \quad (1)$$

The values of $A$ and $\gamma$ and the best fit curve are given in Figure S4b. This established that the target EML thickness of 55 nm is achieved at a spin speed of approximately 1600±100 rpm.

**Variable Angle Spectroscopic Ellipsometry**

Films for study using Variable-angle Spectroscopic Ellipsometry (VASE) were spin coated on Eagle-XG glass and silicon substrates. VASE measurements were performed with a Wollam M2000 ellipsometer (ESM 300, EC-400, M-2000XI). Data was analysed in the Wollam's CompleteEASE software. Optical constants and film thicknesses were obtained by fitting the ellipsometry angles using several empirical models (Cauchy, B-spline, generic Gaussian oscillators) [28].

**Atomic Force Microscopy**

Two unencapsulated solution processed films of PEDOT:PSS:PFI on silicon substrates were interrogated employing an atomic force microscope (Bruker JPK NanoWizard® 4 BioAFM) operating in non-contact mode. The AFM investigation was conducted on an air floated optical table to decouple from environmental acoustic interference. A line-by-line scan of a 10x10 $\mu m^2$ area was conducted (512x512 px). Data processing was handled via the open-source Gwyddion software [46]. Image manipulations included 2D FFT high-frequency noise filtering, image mean slope correction and 4th-order background polynomial subtraction.

**Emitter orientation characterisation**

Angle-resolved photoluminescence (ARPL) measurements were performed on encapsulated EML/HIL/glass samples. ARPL is used to determine the average orientation of emissive transition dipoles from an ensemble of molecules in a thin film [40], [47]. The key quantity extracted from the ARPL analysis is the average transition dipole moment (TDM) orientation factor, defined as the average of squared cosines of the dipole angles $\vartheta$ with respect to the substrate normal:

$$a_{ARPL} = \langle\cos^2 \vartheta\rangle. \quad (2)$$

The ARPL data are fitted by minimising the squared residual between the measured profiles and a linear combination of simulated horizontal ($I_{||}$) and vertical ($I_{\perp}$) dipole power dissipation density spectral profiles:

$$I(\lambda, \vartheta) = (1 - a_{ARPL})\, I_{||}(\lambda, \vartheta) + a_{ARPL}\; I_{\perp}(\lambda, \vartheta). \quad (3)$$

These profiles are computed using transfer matrix method (TMM) calculations for the optical thin-film stack, bounded by semi-infinite air and substrate media, using the spectrally resolved complex refractive indices for all layers [47], [48]. The resulting orientation factor for the mCPCN:SpiroAC-TRZ EML is shown in Figure S5b.

**OLED fabrication and characterization**

Glass substrates with 90 nm-thick patterned ITO electrodes (sheet resistance 20 ohm/sq, Xin Yan Technology) were cleaned by wiping with lint-free tissues rinsed in isopropanol (IPA), followed by ultrasonication in deionised water, acetone, and IPA in series for 10 minutes each, and subsequently treated with UV-ozone for 2–3 minutes. The PEDOT:PSS:PFI HIL was then spin-coated in air under laminar flow for 25 s at 4000 rpm from a dispersion in IPA and deionised water (DI $H_2O$) with a PEDOT:PSS:PFI mass ratio of 1:6:14. The IPA/DI $H_2O$/PEDOT:PSS:PFI mass ratio of 82.2/17.05/0.75 (81.7/16.8/1.5) wt% in the dispersion yielded 25 (50) nm thick films. The substrates were annealed at 130 °C for 10 minutes. For the SpiroAC-TRZ devices, 50 nm thick PEDOT:PSS:PFI films were exposed to an RF $O_2$ plasma at a chamber pressure of 0.2 Pa for 20 s (PlasmaFlecto 10, Plasma Technology) to increase wetting by aromatic solvents, for the remaining devices, untreated 25 nm thick PEDOT:PSS:PFI films were used. Substrates were transferred into a $N_2$-filled glovebox (Glovebox Systemtechnik), where the various EMLs were deposited by dropwise applying 20 mg mL$^{-1}$ chlorobenzene solutions on the anode area with a 1-2 mm lateral margin and spin-coating at 1600-1700 rpm. Next, the samples were annealed at 85 °C for 30 minutes. EML compositions for the different device types are given in the relevant text of the Results section. The remaining layers — TmPyPB (25-40 nm), LiF (1 nm), and Al (100 nm) — were thermally evaporated under high vacuum (<$10^{-7}$ mbar, Angstrom Engineering) at deposition rates of 0.6, 0.1, and 1.0 Å s$^{-1}$, respectively. Shadow masks were used to define pixel areas of 2 × 2 mm². Devices were encapsulated by applying UV-curable epoxy around the device perimeter, placing encapsulation glasses on top, and curing under UV light for 2 minutes with the active pixel area masked from UV exposure.

The completed devices were characterised using a calibrated setup developed in-house, capable of measuring *J*–*V*–*L* characteristics and angle-resolved electroluminescence spectra (AREL) [47], from which luminance and EQE values were extracted.

**Acknowledgement**

This project has been partly funded by the European Union Horizon 2020 research and innovation programme under grant agreement No. 101073045 (TADFsolutions), the EPSRC grant EP/X026175/1, and instrument funding by the Deutsche Forschungsgemeinschaft in cooperation with the Ministerium für Kunst und Wissenschaft of North Rhine-Westphalia (469988234, INST 216/1120-1 FUGG). M.C.G. acknowledges funding from the Alexander von Humboldt Foundation (Humboldt Professorship) and the European Research Council under the Europeans Union's Horizon / ERC Advanced Grant agreement No. 101097878 (HyAngle). E.Z.-C. thanks the Engineering and Physical Sciences Research Council for support (EP/Z535291/1 and EP/W007517/1).

**Conflicts of Interest**

The authors declare no conflicts of interest.

**Supporting Information**

Additional supporting information can be found online in the Supporting Information section.

**Data Availability Statement**

The data that support the findings of this study are openly available in the University of St Andrews Research Portal at [DOI assigned on acceptance].

**Author contributions**

**Džiugas Litvinas**: Conceptualization, Investigation, Writing – original draft. **Francisco Tenopala-Carmona**: Resources, Supervision, Writing – review & editing. **Mahni Fatahi**: Resources, Writing – review & editing. **Anna Popczyk**: Investigation, Methodology. **Eli Zysman-Colman**: Supervision, Writing – review & editing. **Malte C. Gather**: Conceptualization, Supervision, Writing – review & editing.

# Sky-blue Solution-processed TADF and Hyperfluorescent OLEDs via a High-Work-Function PEDOT:PSS:PFI Hole Injection Layer

Džiugas Litvinas [1], Francisco Tenopala-Carmona [1], Mahni Fatahi [2], Anna Popczyk [1], Eli Zysman-Colman* [2], Malte C. Gather* [1,3]

## Supporting Information

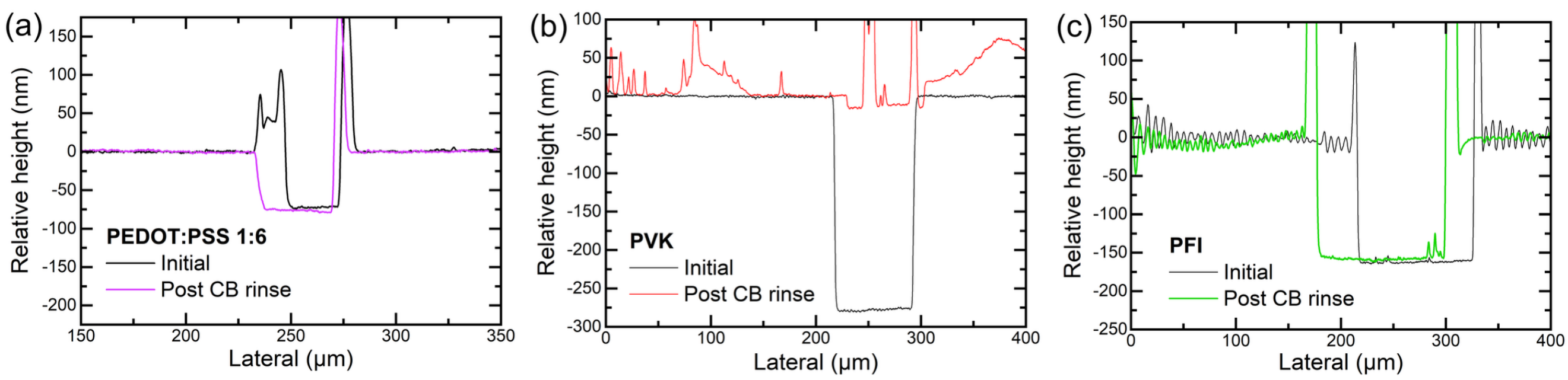

Figure S1. Thickness profiles across scratches in films of (a) PEDOT:PSS, (b) PVK and (c) PFI, before and after rinsing with chlorobenzene (CB).

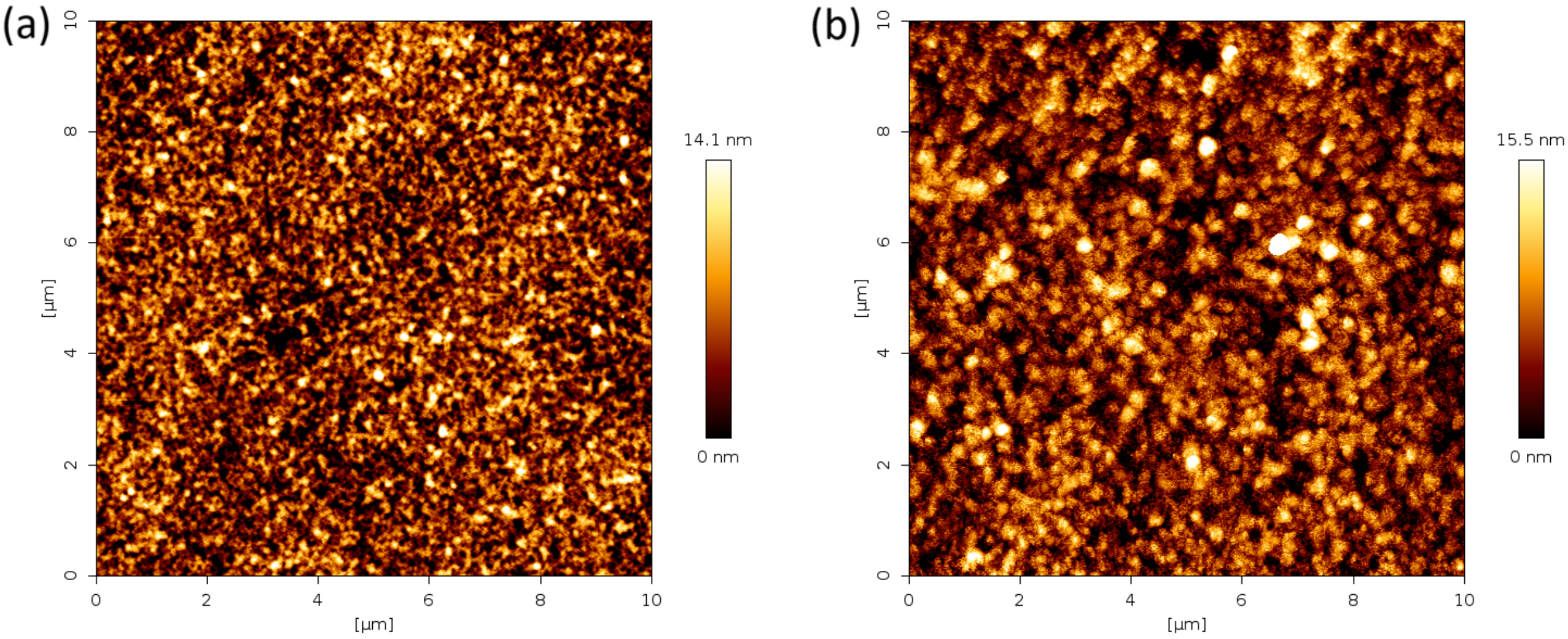

Figure S2. AFM topographic images (non-contact mode) of a PEDOT:PSS:PFI layer (a) prior to treatment, (b) after RF $O_2$ plasma treatment.

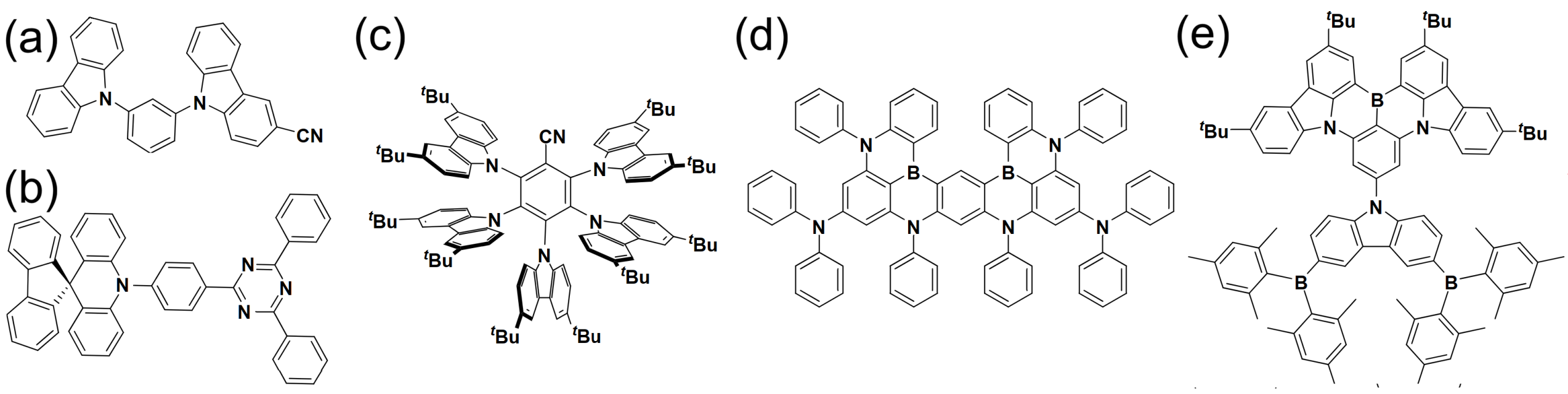

Figure S3. Molecular structures of the solution-processed EML materials used in this work: (a) the host mCPCN; (b) the TADF emitter SpiroAC-TRZ; (c) the stable TADF sensitizer 5TCzBN; (d) the benchmark MR-TADF emitter ν-DABNA; (e) the MR-TADF compound TCzBN-BMes.

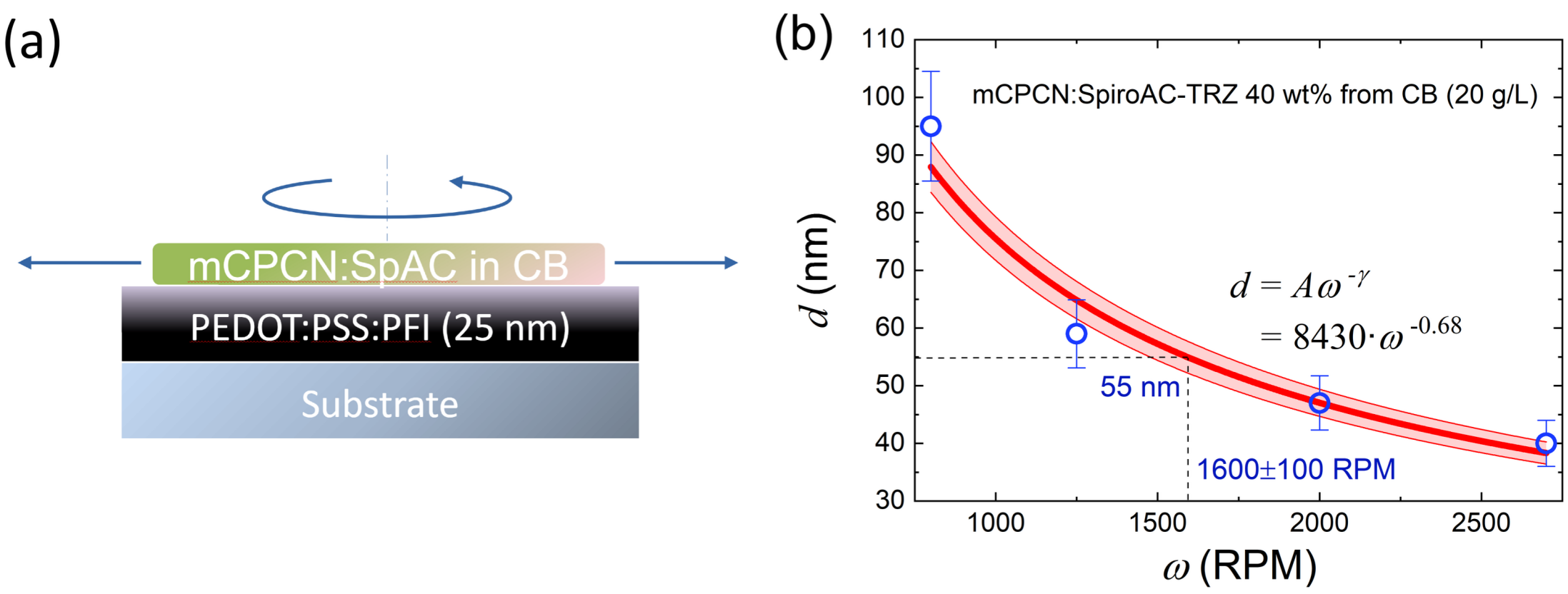

Figure S4. (a) Schematic illustration of the spin-coating geometry used for EML thickness calibration. (b) Spin-coated layer thickness calibration curve for mCPCN:SpiroAC-TRZ (40 wt%) on top of the PEDOT:PSS:PFI HIL (after $O_2$ plasma treatment).

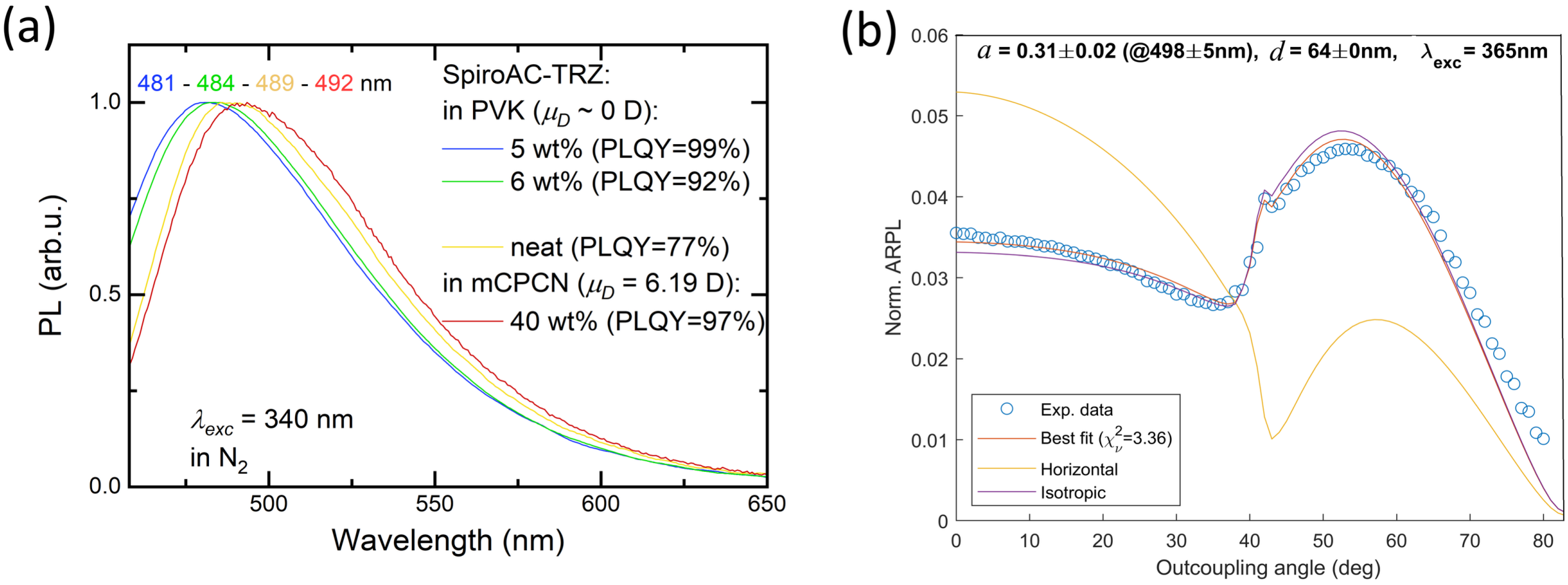

Figure S5. (a) Normalised photoluminescence spectra and PLQY values for SpiroAC-TRZ in a range of host environments: neat film, doped into the low-polarity host PVK (dipole moment $\mu_D \approx 0$ D) at 5 and 6 wt%, and doped into the polar host mCPCN ($\mu_D = 6.19$ D) at 40 wt%. Spectra were collected under nitrogen atmosphere with excitation at 340 nm. The substantially higher PLQY achieved in mCPCN at 40 wt% relative to PVK at low doping concentrations motivates the choice of mCPCN as the host material in this work. (b) Angle-resolved photoluminescence (ARPL) data and best-fit simulated curve for a mCPCN:SpiroAC-TRZ (40 wt%) EML spin-coated on top of a 25 nm thick PEDOT:PSS:PFI HIL, yielding an anisotropy parameter a = 0.31 ± 0.02, corresponding to a largely isotropic emitter orientation. Simulated ARPL profiles for the fully horizontal and perfectly isotropic case are shown for comparison.

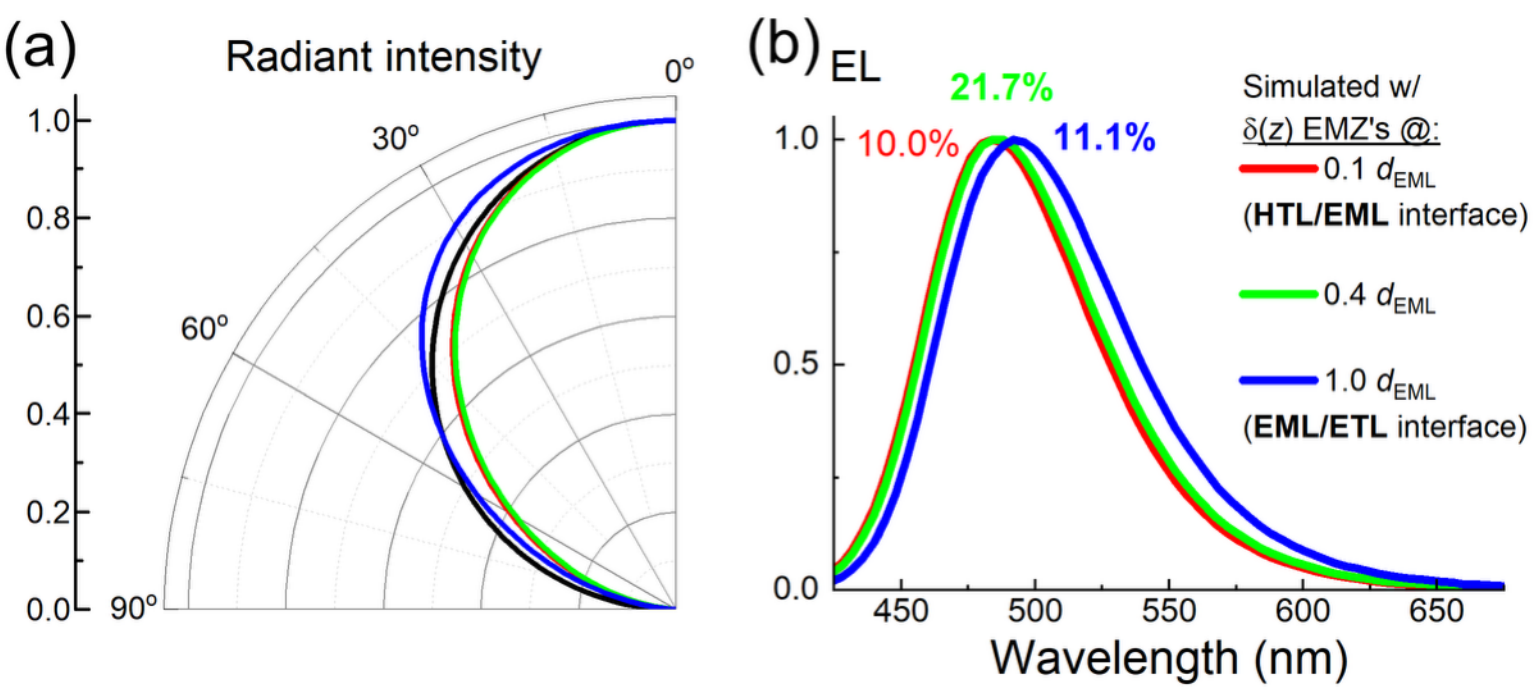

Figure S6. Simulated angular electroluminescence (AREL) profiles (a) and spectra (b) of transfer matrix optical calculations for the mCPCN:5TCzBN on PEDOT:PSS:PFI OLED stack (Figure 4b, main text), assuming infinitesimally thin emission zones at three limiting positions: at the HIL/EML interface (red), near the centre of the EML at the position of optimal outcoupling (green), and at the EML/ETL interface (blue). The EQE values annotated on the spectra correspond to the simulated efficiencies for each emission zone position.

**FRET Efficiency Calculation**

The Förster resonant energy transfer radius $R_{FRET}$ and efficiency $\eta_{FRET}$ were computed using the expressions [49], [50], [51]:

$$R_{FRET} = ([9[\ln 10]\kappa^2 \Phi_D J\,]/[128\pi^5 N_A])^{1/6}, \tag{S1}$$

$$J = \int PL_D(\lambda)\epsilon_A(\lambda)\ [\lambda\ /\ n(\lambda)]^4\ \mathrm{d}\lambda, \tag{S2}$$

$$\eta_{FRET} = 1\ /\ (1 + [\langle r\rangle_{D\text{-}A}\ /\ R_{FRET}]^6), \tag{S3}$$

where $\kappa^2$ = 2/3 is the orientational factor for interacting dipoles having no orientation preference (isotropic), $\Phi_D$ = 80% is the photoluminescence quantum yield of the donor (5TCzBN in toluene [41]), $N_A$ = 6.0221408 × $10^{23}$ $mol^{-1}$ is Avogadro's number, $n(\lambda)$ is the refractive index spectrum of the host-donor matrix, and $J$ is the spectral overlap integral between the acceptor molar absorption coefficient $\epsilon_A(\lambda)$ and the area-normalised donor emission $PL_D(\lambda)$ spectra (Figure 5a, b). Since in the context of emissive layers the molecules are in solid state, we can assume an ordered arrangement of the host and donor molecules. Furthermore, let the donor molecules by themselves be arranged in a pseudo-cubic lattice. The donor lattice constant $a$ is estimated from the donor weight concentration in the host matrix $[\mathrm{5TCzBN}]_{\mathrm{wt\%}}$, and by assuming a cubic packing with a lattice constant of the mCPCN host $a_{\mathrm{mCPCN}}$ = 1.35 nm (taken from [27]). Thus,

$$a = a_{\mathrm{mCPCN}}\ /([\mathrm{5TCzBN}]_{\mathrm{wt\%}}\ M_{\mathrm{5TCzBN}}\ /\ M_{\mathrm{mCPCN}})^{1/3}, \tag{S4}$$

where $M_{\mathrm{mCPCN}}$ = 433.5 g/mol ($M_{\mathrm{5TCzBN}}$ = 1490.1 g/mol) is the molar weight of the host (donor). More specifically, for a $[\mathrm{5TCzBN}]_{\mathrm{wt\%}}$ = 10 wt% (30 wt%), the resulting $a$ = 4.39 nm (3.04 nm). The following assumptions are also made: 1) the acceptor concentration is far smaller than the donor one, 2) all donor sites are excited (not valid for low exciton densities).

Moving on, two reasoning approaches for estimating the average donor-acceptor distance $\langle r\rangle_{D\text{-}A}$ are discussed. We start with the statistical quasi-physical model treatment. The spatial probability density of acceptor occurrence is modelled by dipole-dipole interaction: $p(\mathbf{r}) = p_0\ \Sigma_i\ 1/\ [1{+}b\times(\mathbf{r}{-}\mathbf{r}_i)^6]$, with $\mathbf{r}$ and $\mathbf{r}_i$ being the radius vectors of the acceptor and the i-th neighbouring donor, respectively, and $b$ the interaction strength; only the terms of the nearest vertex and its 26 closest neighbours are accounted for (Figure S7a). The normalization factor $p_0$ is obtained as:

$$\begin{aligned} p_0 &= 1\ /\ [\textstyle\int_0^{a/2} \mathrm{d}x \int_0^{a/2} \mathrm{d}y \int_0^{a/2} \Sigma_i\ 1/[1 + b\times(\mathbf{r}-\mathbf{r}_i)^6]\ \mathrm{d}z] \\ &= a^{-3}\ /\ [\textstyle\int_0^{1/2} \mathrm{d}x' \int_0^{1/2} \mathrm{d}y' \int_0^{1/2} \Sigma_i\ 1/[1 + b_0\times(\mathbf{r}'-\mathbf{r}_i')^6]\ \mathrm{d}z']. \end{aligned} \tag{S5}$$

Here we normalize the variable and parameter scale: $x = a\,x'$, $b = a^{-6}\,b_0$. Note that $\langle r\rangle_{D\text{-}A}$ is then the average distance from an acceptor to the nearest donor. For simplicity's sake, we calculate the weighted average distance from a donor vertex in one octant of the cubic unit cell:

$$\langle r\rangle_{D\text{-}A} = \int_0^{a/2} \mathrm{d}x \int_0^{a/2} \mathrm{d}y \int_0^{a/2} p(\mathbf{r})\ |\mathbf{r}|\ \mathrm{d}z. \tag{S6}$$

The integrals (S5) and (S6) were calculated numerically in Python for different $b_0$ values. The results are plotted in Figure S8. Minimum and maximum average distance values are observed for $b_0$ = 4.978(7) × $10^{-2}$ and $b_0 \to \infty$, respectively. To have a conservative estimate for FRET efficiency, the greatest possible $\langle r\rangle_{D\text{-}A}$ value ought to be used. Thus, leading to $\langle r\rangle_{D\text{-}A}^{dipole\text{-}model}$ = 0.5584(7) × $a$.

The second approach is far simpler, where we assume that the acceptor molecules are on average as far away from all of the cubic donor lattice vertices as possible, meaning that the acceptor is situated at the centre of the elementary lattice cube (Figure S7b). Thus, $\langle r\rangle_{D\text{-}A}^{centre} = \sqrt{3}/2 \times a$ = 0.8660(3) × $a$. Seeing as how the latter estimate is more conservative, it can be taken as the main case for our FRET calculations.

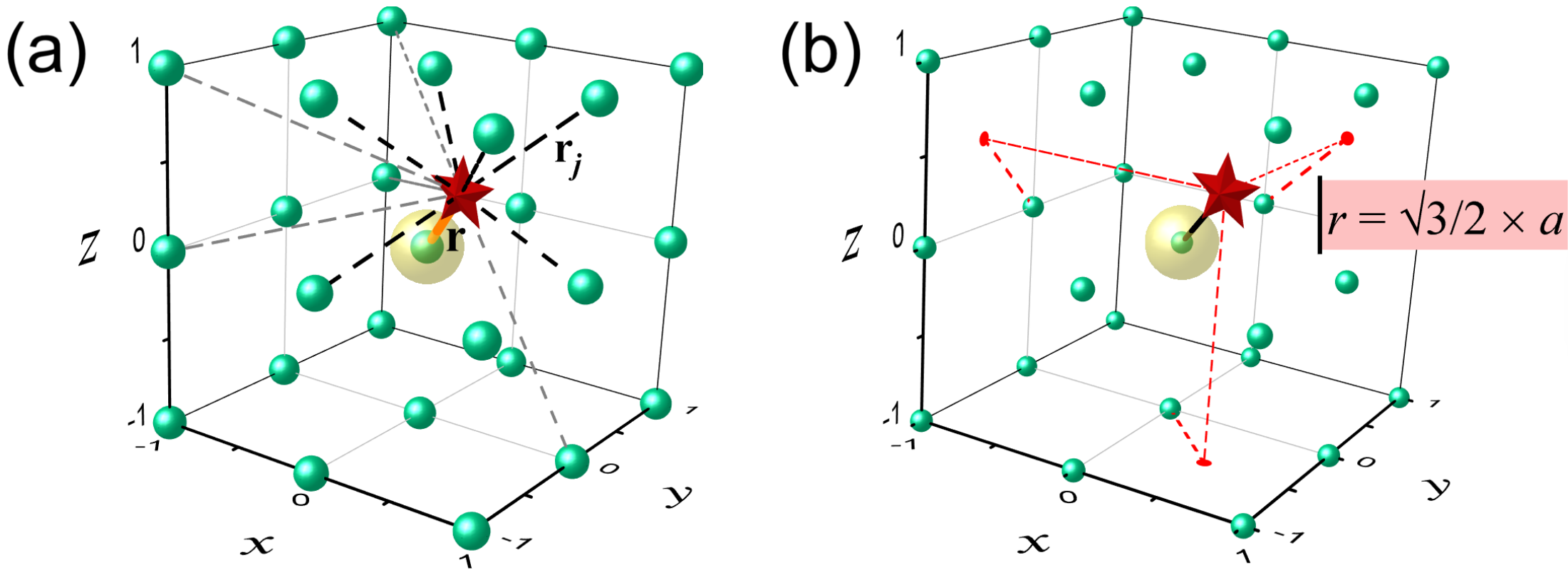

Figure S7. Simplified representations of the relative geometries between the effective donor lattice (blueish-green spheres) and the acceptor molecule (red star), used in $\langle r \rangle_{D\text{-}A}$ estimation, assuming the donor site closest to the acceptor to be excited. The two subplots depict either (a) the interaction distances between lattice vertices and the acceptor dipoles, or (b) the greatest overall distance possible between any cubic lattice donor and the acceptor.

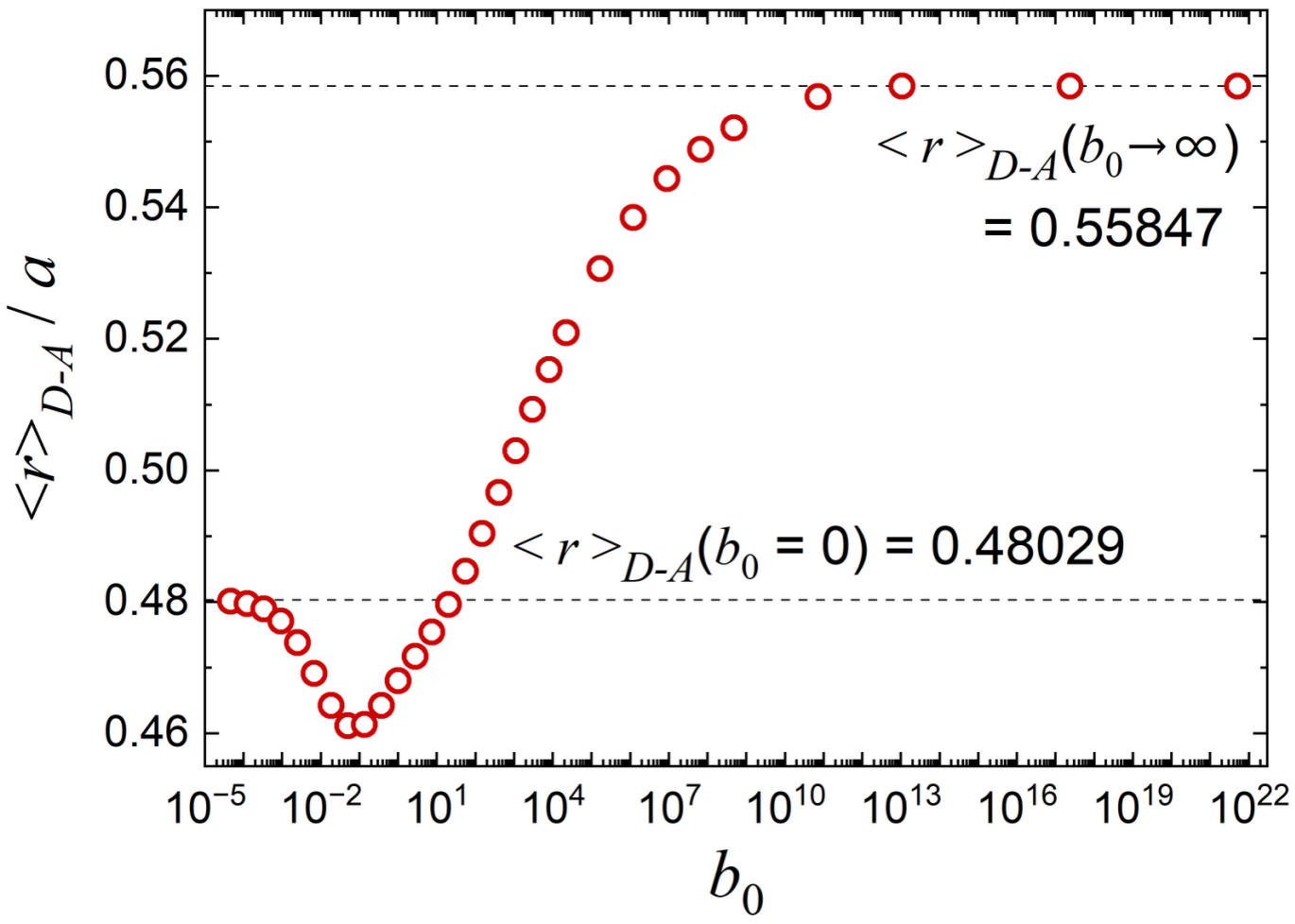

Figure S8. Simulated average donor-acceptor distances $\langle r \rangle_{D\text{-}A}^{dipole\text{-}model}$ for a cubic donor lattice assuming a dipole-dipole interaction-based acceptor spatial probability density. The x-axis parameter is the arbitrary unitless interaction strength $b_0$.